\documentclass[12pt]{article}
\usepackage{amssymb}

\renewcommand{\thefootnote}{\fnsymbol{footnote}}

\begin{document}

\title{
\begin{flushright}
\ \\*[-80pt] 
\begin{minipage}{0.2\linewidth}
\normalsize
arXiv:0801.0998 \\
YITP-07-98 \\
TU-787 \\
KUNS-2118 \\*[50pt]
\end{minipage}
\end{flushright}
{\Large \bf 
Dynamically sequestered F-term uplifting \\
in extra dimension
\\*[20pt]}}

\author{
\centerline{
Hiroyuki~Abe$^{1,}$\footnote{
E-mail address: abe@yukawa.kyoto-u.ac.jp}, \ 
Tetsutaro~Higaki$^{2,}$\footnote{
E-mail address: tetsu@tuhep.phys.tohoku.ac.jp}, \ 
Tatsuo~Kobayashi$^{3,}$\footnote{
E-mail address: kobayash@gauge.scphys.kyoto-u.ac.jp} \ and \ 
Yuji~Omura$^{4,}$\footnote{E-mail address: omura@scphys.kyoto-u.ac.jp}} 
\\*[20pt]
\centerline{
\begin{minipage}{\linewidth}
\begin{center}
$^1${\it \normalsize 
Yukawa Institute for Theoretical Physics, Kyoto University, 
Kyoto 606-8502, Japan} \\
$^2${\it \normalsize 
Department of Physics, Tohoku University, 
Sendai 980-8578, Japan} \\
$^3${\it \normalsize 
Department of Physics, Kyoto University, 
Kyoto 606-8502, Japan} \\
$^3${\it \normalsize 
Department of Physics, Kyoto University, 
Kyoto 606-8501, Japan} 
\end{center}
\end{minipage}}
\\*[50pt]}

\date{
\centerline{\small \bf Abstract}
\begin{minipage}{0.9\linewidth}
\medskip 
\medskip 
\small
We study concretely several issues altogether, moduli stabilization, 
the dynamical supersymmetry (SUSY) breaking, the uplifting of 
SUSY anti-de Sitter (AdS) vacuum and the sequestering of hidden 
sector, in a simple supergravity model with a single 
extra dimension. The sequestering is achieved dynamically by a 
wavefunction localization in extra dimension. The expressions 
for the visible sector soft SUSY breaking terms as well as the 
hidden sector potential are shown explicitly in our model. 
We find that the tree-level soft scalar mass and the A-term can 
be suppressed at a SUSY breaking Minkowski minimum where the 
radius modulus is stabilized, while gaugino masses would be 
a mirage type. 
\end{minipage}
}

\begin{titlepage}
\maketitle
\thispagestyle{empty}
\end{titlepage}


\renewcommand{\thefootnote}{\arabic{footnote}}
\setcounter{footnote}{0}

\section{Introduction}
Supersymmetric extensions of the standard model (SM) are promising 
candidates for the physics around a TeV scale. The supersymmetry (SUSY), 
which is predicted by a known consistent theory of the quantum gravity, 
i.e., superstring theory, can protect the electroweak scale 
$M_{EW} \sim 10^2$ GeV against radiative corrections of the order of 
the Planck scale $M_{Pl} \sim 10^{19}$ GeV, even with soft SUSY
breaking. That means, 
the unification of the gravity and the SM would be possible within 
such framework without violating the mass scale of the SM. Moreover, 
the minimal supersymmetric standard model (MSSM) suggests that the 
three gauge couplings in the SM are unified at the grand unification 
theory (GUT) scale, $M_{GUT} \sim 10^{16}$ GeV. 
The lightest SUSY particle (LSP) can be a candidate for the cold dark 
matter, if anything like R-parity forbids LSP decays. 

Because any SUSY particles have not been observed yet, the SUSY must 
be broken above the electroweak scale. The SUSY breaking generically 
introduces flavor violating masses, mixing and couplings (soft SUSY 
breaking terms), which are severely restricted by the flavor changing 
neutral current experiments. If the SUSY breaking effects are 
dominantly mediated from the SUSY breaking (hidden) sector to the 
visible (MSSM) sector by gauge interactions which are flavor blind, 
SUSY flavor violations can be suppressed. However, if the dominant 
SUSY breaking effects are mediated by gravitational interactions 
which are generically flavor dependent, there must exist certain 
mechanism to suppress flavor violations in the visible sector, 
which is sometimes called sequestering. The ultimate situation 
for the sequestering is the case that the hidden sector is 
completely separated from the visible sector by the 
spatiality of extra dimensions~\cite{Randall:1998uk}. 
In this case, the soft SUSY breaking terms 
in the visible sector is generated through the superconformal 
anomaly~\cite{Randall:1998uk,Giudice:1998xp}. 
However, due to the ultra-violet (UV) insensitive nature of 
anomaly mediation, the low energy behavior of MSSM soft terms 
are completely determined, resulting in tachyonic slepton masses. 

On the one hand, if we consider superstring theory (or its 
effective supergravity theory) as a UV completion of the MSSM (or any 
SUSY SM), there generically exist moduli superfields in the 
four-dimensional (4D) effective theory. Vacuum expectation 
values (VEVs) of moduli correspond to sizes and shapes of 
extra dimensions and determine quantities in the 4D effective 
theory such as $M_{Pl}$, gauge and Yukawa couplings. 
Naively, the moduli are flat directions of the potential, 
but must be stabilized by some nontrivial effects such as fluxes 
and/or nonperturbative effects. The moduli stabilization is also 
quite relevant to soft SUSY breaking terms because moduli 
multiplets generically couple to the visible sector in the 
effective theory, and the auxiliary components of the moduli 
multiplets are also determined by the potential stabilizing 
the moduli themselves in supergravity. 

It has been recognized that the realization of a SUSY breaking 
vacuum with an (almost) vanishing vacuum energy, which is 
required by the observations, is quite difficult in the 
conventional moduli stabilization schemes in supergravity theory. 
This is mainly because the moduli potential in the effective 
supergravity theory, in general, prefers SUSY preserving anti-de 
Sitter (AdS) vacua with a negative vacuum energy. 
Recently, a systematic way for realizing a 
SUSY breaking Minkowski minimum in a controllable manner 
was proposed, which we call the Kachru-Kallosh-Linde-Trivedi 
(KKLT) scenario~\cite{Kachru:2003aw}. 
In this scenario, we uplift the above mentioned SUSY AdS 
vacuum to a Minkowski minimum by a SUSY breaking vacuum energy 
generated in the so-called uplifting sector, which is assumed 
to be well sequestered from the light moduli as well as 
the visible sector. In the original KKLT scenario, the 
uplifting sector is formed by an anti D3-brane located at 
the tip of the warped throat in extra dimensions. 

Because of the geometrical structure, the SUSY breaking anti-brane 
can be sequestered from the light moduli and the visible sector 
(on D7 branes). 
Soft SUSY breaking terms are calculated~\cite{Choi:2004sx,Choi:2005ge} 
in the KKLT model. It was found that the tree level (light) modulus 
mediation is generically comparable to the anomaly mediation, resulting 
in the so-called mirage mediation~\cite{Choi:2005uz} 
and leading to phenomenologically interesting 
aspects ~\cite{Endo:2005uy,Choi:2005hd}. 
Then, this model could be restricted from the viewpoint 
of flavor violation, if the light-modulus couplings to 
the visible sector are flavor dependent. The flavor dependence 
of the modulus couplings might be determined by the mechanism of 
generating Yukawa hierarchies for quarks and leptons. 
However, it is still quite a challenging issue to realize successful 
Yukawa structures within superstring models. 

The anti-brane breaks SUSY explicitly in the effective supergravity 
theory. 
We can modify the KKLT scenario such that the uplifting energy is 
supplied by nonvanishing F-terms~\cite{Saltman:2004sn,Dudas:2006gr,
Abe:2006xp,Abe:2007yb} and/or D-terms~\cite{Burgess:2003ic,Choi:2006bh} 
in the dynamical SUSY breaking (hidden) sector\footnote{
Note that we call the dominant SUSY breaking sector as the 
hidden sector, but the moduli sector is not included 
in the hidden sector in our terminology.} 
such as the O'Raifeartaigh model~\cite{O'Raifeartaigh:1975pr} 
and the Intriligator-Seiberg-Shih (ISS) 
model~\cite{Intriligator:2006dd}.\footnote{These models have the same 
behavior when heavy modes are integrated out.} 
The former is called as the F-term uplifting, which can realize a low energy 
SUSY models more easily than the latter. In this scenario, 
the soft terms are generically model dependent (see, e.g., 
Ref~\cite{Choi:2007ka} and references therein), due to the possible direct 
couplings between the hidden and the visible sectors. 
If contact terms between the visible and hidden sectors are 
not suppressed, soft SUSY breaking scalar masses in the 
visible sector would of ${\cal O}(m_{3/2})$.
However, scalar masses would be suppressed if 
the hidden sector is sequestered anyhow from the visible sector.
At any rate, these scalar masses depend on explicit 
forms of couplings between the hidden and visible sectors.
Thus, we should
identify (or construct) the hidden sector explicitly 
and clarify explicit couplings between the visible and
hidden sectors in order to calculate soft SUSY 
breaking terms in the F-term uplifting scenario.
Therefore, our purposes of this paper are to study 
the F-term uplifting scenario in a simple model, where 
direct couplings between the visible and hidden sectors 
are clarified explicitly, and to analyze soft SUSY breaking terms 
including consideration on sequestering.

As we find from the above overview, a whole phenomenological study 
of the moduli stabilization, spontaneous SUSY breaking, 
uplifting and sequestering 
has not been done explicitly. 
This is simply because effective supergravity theories of 
superstring models are complicated enough 
or difficult to be derived, 
preventing a systematic study. In this paper, we study concretely 
these issues altogether in a simple supergravity model 
with a single extra dimension, i.e., five-dimensional (5D)
supergravity. 
By orbifolding 
the extra dimension and considering boundary induced potential terms, 
all the above mechanisms for stabilizing moduli, breaking SUSY and 
uplifting AdS vacua would be realized, and the soft SUSY breaking 
terms can be calculated explicitly, with which we can study the 
feature of sequestering in our model. Moreover it is known that a 
certain class of realistic Yukawa matrices for quarks and leptons 
can be obtained by the wavefunction localization (see, e.g., 
Refs.~\cite{Choi:2003di,Abe:2004tq} and references therein) 
as a solution of the equation of motion in 5D 
supergravity~\cite{Gherghetta:2000qt}. 
Here we use such a localization mechanism to sequester the hidden 
sector from the visible sector. Although our model is not directly 
related to a certain string model known at present, we believe that 
studying such a simple setup helps us to understand basic and important 
natures underlying the above issues in effective supergravity theory. 
Of course, our supergravity model itself could be a candidate 
for the physics beyond the MSSM. 

The following sections are organized as follows. 
In Sec.~\ref{sec:model}, we construct a 5D model for dynamically 
sequestering a SUSY breaking (F-term uplifting) sector by the 
wavefunction localization in the extra dimension. We also derive 
the 4D effective Lagrangian for our model. 
In Sec.~\ref{sec:soft}, we compute the K\"ahler potential and the 
superpotential for the radius modulus $T$ and the hidden sector 
field $X$, which are relevant to the dynamical SUSY breaking and 
the uplifting. We also show the resulting soft SUSY breaking 
terms in terms of the auxiliary components $F^T$ and $F^X$. 
Then, we discuss about the nature of SUSY breaking, F-term 
uplifting and sequestering in our model based on them. 
We summarize this paper in Sec.~\ref{sec:summary}. 
In Appendix~\ref{app:ftu}, the basic structure of F-term uplifting 
is briefly reviewed as well as the original KKLT model.

\section{Quasi-localized visible and hidden sectors}
\label{sec:model}

\subsection{5D model}
We consider a model in which the SUSY breaking sector can be 
sequestered from the visible sector by an equation of 
motion in extra dimension. The 5D supergravity compactified 
on an orbifold provides a simple framework for such a model, 
and we can construct an illustrating example of dynamical 
sequestering, SUSY breaking and F-term uplifting. 

For such a purpose, we start from the 5D off-shell (conformal) 
supergravity on an orbifold $S^1/Z_2$~\cite{Kugo:2002js} 
with the radius $R$. For simplicity and concreteness, 
we choose the simplest bulk supergravity characterized 
by a single $Z_2$-odd vector multiplet (graviphoton) and 
a single compensator hypermultiplet in addition to 
matter multiplets. Because of that, we have a single (radius) 
modulus $T$ in the 4D effective theory, whose VEV is 
related to the orbifold radius as $\langle T \rangle=\pi R$, 
and the target manifold of the hypermultiplet is characterized 
by $USp(2,2n_H)/USp(2) \times USp(2n_H)$, 
where $n_H$ is the number of the matter hypermultiplets. 
The extensions to the case with more odd vector multiplets 
(i.e., more moduli) and/or with more compensators are 
straightforward.\footnote{For example, the power $2/3$ of 
the integrand in Eq.(\ref{eq:omega_bulk}) is different in 
the case with two compensators, from the viewpoint of 
the 4D effective theory.} 

As for matter multiplets, we embed visible sector (MSSM) 
chiral multiplets $Q^I$ into 5D hypermultiplets 
${\cal H}^I=({\cal Q}^I,{\cal Q}_c^I)$, 
where $Q^I$ is the zero-mode of ${\cal Q}^I$, and the index 
$I$ runs over all the quarks, leptons and Higgs fields. 
The hidden sector chiral multiplet $X$, which is responsible 
for the dynamical SUSY breaking and for the uplifting, is also 
assigned to a 5D hypermultiplet ${\cal H}^X=({\cal X},{\cal X}_c)$ 
where $X$ is the zero-mode of ${\cal X}$. 
Zero modes of  chiral multiplet 
partners ${\cal Q}_c^I$ and ${\cal X}_c$ in these hypermultiplets 
are projected out by orbifolding. 

We gauge $U(1)$ isometries of the compensator, visible (MSSM) 
and hidden hypermultiplets by the graviphoton with corresponding 
charges $k$, $c_I$ and $c_X$, 
respectively~\cite{Gherghetta:2000qt,Falkowski:2000er,Fujita:2001bd}. 
Because the graviphoton has an odd $Z_2$-parity under the 
orbifold projection, the gauge coupling should change its sign 
across the orbifold fixed points located at $y=0,\pi R$. 
This can be achieved by accompanying a periodic sign function 
$\epsilon(y)$ with the gauge coupling.\footnote{In supergravity, 
this can be achieved by the so-called four-form mechanism~\cite{
Bergshoeff:2000zn,Fujita:2001bd}.} 
Then, these gauging procedures generate kink-type masses 
$\epsilon(y) \hat{k}$, 
$\epsilon(y) \hat{c}_I$ and $\epsilon(y) \hat{c}_X$ for the gravitino, 
visible and hidden hyperinos, respectively, as well as a bulk 
negative cosmological constant scaled by $\hat{k}$, where 
$\hat{k}=kM$, $\hat{c}_I=c_IM$, $\hat{c}_X=c_XM$ and $M$ 
is the VEV of graviscalar (a scalar component of graviphoton 
multiplet). Thus, the equations of motion in the orbifold 
segment generate exponential profiles $e^{-\hat{k}y}$, 
$e^{-\hat{c}_Iy}$ and $e^{-\hat{c}_Xy}$ for the 
wavefunctions~\cite{Gherghetta:2000qt} 
of the gravitino, visible and hidden hyperinos, respectively, 
in the orbifold slice of AdS$_5$ warped background 
geometry\footnote{The boundary tension terms which balance 
with the bulk cosmological constant is automatically 
supplied~\cite{Fujita:2001bd} by the four-form mechanism 
mentioned in the previous footnote.} $ds^2=e^{-2\hat{k}y}dx^2-dy^2$. 

The bulk supergravity has an $N=2$ SUSY (eight supercharges), 
and does not allow Yukawa interactions between hypermultiplets. 
To be phenomenologically viable, we introduce superpotential 
terms at the fixed point including the Yukawa interaction 
between quarks/leptons and the Higgs bosons. In addition, 
some superpotential terms for the hidden sector field $X$ 
are necessary for triggering a dynamical SUSY breaking, 
generating a nonvanishing value of $F^X$, which would uplift the 
negative vacuum energy of the modulus sector. 
Actually, the fixed points respect only $N=1$ SUSY 
(four supercharges) which survives under the orbifolding, 
where we can write any K\"ahler/superpotential terms and 
gauge kinetic functions if the other symmetries such as MSSM 
gauge symmetries do not forbid them.

\subsection{4D effective theory}
To study the nature of SUSY breaking, uplifting and sequestering, 
a 4D effective theory of our 5D model is desired. Especially, the 
information of contact interactions between $Q^I$ and $X$ in the 
K\"ahler potential is important in such 4D effective theory, in 
order to analyze the sequestering structure. For this end, we adopt 
the off-shell dimensional reduction method proposed by 
Refs.~\cite{Correia:2006pj,Abe:2006eg}, 
which is based on an $N=1$ superspace description of 5D conformal 
supergravity on an orbifold~\cite{Paccetti:2004ri} 
and developed in subsequent works~\cite{Abe:2005ac}. 
This method provides us a way for deriving the 4D off-shell 
effective action directly from the 5D off-shell supergravity action 
with generic boundary terms, respecting the $N=1$ off-shell structure. 
The procedure is as follows. We start from the $N=1$ off-shell 
description of 5D action. After some gauge transformation, we drop 
kinetic terms for $Z_2$-odd multiplets which are negligible at low 
energy. Then, these multiplets play a role of Lagrange multiplier 
and their equations of motion extract zero-modes from the $Z_2$-even 
multiplets. 

After these steps, we find the 4D effective Lagrangian 
of our 5D model in the $N=1$ superspace description, 
\begin{eqnarray}
{\cal L} &=& 
\int d^4 \theta\, |C|^2 \Omega
+\left\{ \int d^2 \theta\, 
\left( f_a W^{a \alpha} W^a_{\ \alpha} 
+C^3 W \right) +\textrm{h.c.} \right\}, 
\nonumber
\end{eqnarray}
where $C$ is the compensator chiral multiplet 
in the 4D $N=1$ supergravity. 
The K\"ahler potential $K=-3 \ln (-\Omega/3)$, 
the superpotential $W$ and 
the gauge kinetic function $f_a$ are give by 
\begin{eqnarray}
\Omega &=& 
\Omega^{({\rm bulk})}
+\sum_{\vartheta=0,\pi} 
e^{-\frac{\vartheta}{\pi} k(T+\bar{T})} 
\Omega^{(\vartheta)} 
\ = \ -3e^{-K/3}, 
\nonumber \\
W &=& 
\sum_{\vartheta=0,\pi} 
e^{-3\frac{\vartheta}{\pi} kT} 
W^{(\vartheta)}+W^{({\rm np})}, 
\nonumber \\
f_a &=& f_a^{({\rm bulk})}
+\sum_{\vartheta=0,\pi} f_a^{(\vartheta)}, 
\nonumber
\end{eqnarray}
and contributions from the bulk are found as 
\begin{eqnarray}
\Omega^{({\rm bulk})} &=& 
-3 \int_0^{{\rm Re}\,T} dt\, e^{-2kt} 
\left( 1 
-e^{-2 \left( c_X-\frac{3}{2}k \right)t}|X|^2 
-\sum_I e^{-2 \left( c_I-\frac{3}{2}k \right)t}|Q^I|^2 
\right)^{\frac{2}{3}}, 
\label{eq:omega_bulk} \\
f_a^{({\rm bulk})} &=& k_a T. 
\nonumber
\end{eqnarray}
The Yang-Mills gauge couplings to the hypermultiplets in 
$\Omega^{({\rm bulk})}$ are omitted to simplify the expression, 
which are irrelevant to the following arguments. 
The nonperturbative effects such as gaugino condensations are 
encoded in the superpotential terms 
\begin{eqnarray}
W^{({\rm np})} &=& 
\sum_{m=1}^{m_{\rm np}} B_m e^{-b_m} 
-\sum_{n=1}^{n_{\rm np}} A_n e^{-a_n T}, 
\nonumber
\end{eqnarray}
where the first constant and the second $T$-dependent term 
would come from the $m_{\rm np}$ boundary and the $n_{\rm np}$ 
bulk (zero-mode) gaugino condensations, respectively. 
Thus the natural orders of the constants are 
\begin{eqnarray}
b_m &\sim& a_n \ = \ {\cal O}(4 \pi^2), \qquad  
B_m \ \sim \ A_n \ = \ {\cal O}(1).
\nonumber
\end{eqnarray}

The other quantities 
\begin{eqnarray}
\Omega^{(\vartheta)} &=& 
\Omega^{(\vartheta)}
(X_\vartheta,\bar{X}_\vartheta,
Q^I_\vartheta,\bar{Q}^I_\vartheta), 
\quad 
W^{(\vartheta)} \ = \ 
W^{(\vartheta)}
(X_\vartheta,Q^I_\vartheta), 
\quad 
f_a^{(\vartheta)} \ = \ 
f_a^{(\vartheta)}
(X_\vartheta), 
\nonumber
\end{eqnarray}
originate from the K\"ahler potential, the superpotential 
and the gauge kinetic function, respectively, induced at 
the 4D fixed point $y=\vartheta R$ ($\vartheta=0,\pi$). 
Within the framework of 5D orbifold supergravity, these 
can be arbitrary functions of the chiral multiplets 
$X_\vartheta$ and/or $Q^I_\vartheta$ originating from the 
bulk hypermultiplets (as well as the boundary own fields 
which are treated implicitly here and hereafter, 
if they are needed), 
\begin{eqnarray}
X_\vartheta &=& e^{-\frac{\vartheta}{\pi} 
\left( c_X-\frac{3}{2}k \right) T }X, \qquad 
Q^I_\vartheta \ = \ e^{-\frac{\vartheta}{\pi} 
\left( c_I-\frac{3}{2}k \right) T }Q^I, 
\nonumber
\end{eqnarray}
for $\vartheta=0$ and $\vartheta=\pi$, respectively. 
In order for the orbifold supergravity theory to be self consistent, 
we consider that all the boundary terms can be treated 
perturbatively, i.e., the field coefficients in 
$\Omega^{(\vartheta)}$, $W^{(\vartheta)}$ and $f_a^{(\vartheta)}$ 
are assumed implicitly to be small compared with 
those originating from the bulk in this paper. 

We consider the 4D effective theory around $Q^I=X=0$, 
and then expand $\Omega$ in powers of $Q^I$ and $X$ as 
\begin{eqnarray}
\Omega &=& 
\hat\Omega_0(T,\bar{T}) 
+Y_{X\bar{X}}(T,\bar{T}) |X|^2 
+Y_{I\bar{J}}(T,\bar{T},X,\bar{X}) Q^I \bar{Q}^{\bar{J}} 
+{\cal O}(Q^4,X^4), 
\nonumber
\end{eqnarray}
where 
\begin{eqnarray}
\hat\Omega_0(T,\bar{T}) &=& 
-3 \alpha \,\frac{1-\beta|e^{-kT}|^2}{2k}, \qquad 
Y_{X\bar{X}}(T,\bar{T}) \ = \ 
\alpha_X\,\frac{1-\beta_X|e^{-(c_X-k/2)T}|^2}{c_X-k/2}, 
\nonumber \\
Y_{I\bar{I}}(T,\bar{T},X,\bar{X}) &=& 
\alpha_I\,\frac{1-\beta_I|e^{-(c_I-k/2)T}|^2}{c_I-k/2}
+\frac{1}{3}\tilde\alpha_I\,
\frac{1-\tilde\beta_I|e^{-(c_I+c_X-2k)T}|^2
}{c_I+c_X-2k}\,|X|^2, 
\label{eq:sswf}
\end{eqnarray}
and the coefficients are determined as 
\begin{eqnarray}
&&
\begin{array}{rclcrcl}
\alpha &=& 1-\frac{2}{3}k \Omega^{(0)}|_0, & & 
\alpha \beta &=& 1+\frac{2}{3}k \Omega^{(\pi)}|_0, 
\\*[5pt]
\alpha_X &=& 1+(c_X-\frac{k}{2}) 
\Omega^{(0)}_{X\bar{X}}|_0, & & 
\alpha_X \beta_X &=& 1-(c_X-\frac{k}{2}) 
\Omega^{(\pi)}_{X\bar{X}}|_0, 
\\*[5pt]
\alpha_I &=& 1+(c_I-\frac{k}{2}) 
\Omega^{(0)}_{I\bar{I}}|_0, & & 
\alpha_I \beta_I &=& 1-(c_I-\frac{k}{2}) 
\Omega^{(\pi)}_{I\bar{I}}|_0, 
\\*[5pt]
\tilde\alpha_I &=& 1+3(c_I+c_X-2k) 
\Omega^{(0)}_{I\bar{I}X\bar{X}}|_0,  & & 
\tilde\alpha_I \tilde\beta_I &=& 
1-3(c_I+c_X-2k) 
\Omega^{(\pi)}_{I\bar{I}X\bar{X}}|_0. 
\end{array}
\nonumber
\end{eqnarray}
Here and hereafter, 
we use the notation that 
$F_{AB \cdots}=
\partial_A \partial_B \cdots F$ 
and $F_{AB \cdots}|_0=
F_{AB \cdots}|_{Q^I=\bar{Q}^{\bar{I}}=X=\bar{X}=0}$ 
for a function $F=F(X,\bar{X},Q^I,\bar{Q}^{\bar{I}})$ 
and indices $A,B,\ldots=(X,\bar{X},I,\bar{I})$. 
In this paper, we assume that the boundary K\"ahler 
potential does not contain flavor mixing, i.e., 
$\Omega_{I\bar{J}}^{(\vartheta)}
|_{Q^I=\bar{Q}^{\bar{I}}=0}=0$ 
for $J \ne I$, for simplicity.\footnote{
We would study a more general case with 
$\Omega_{I\bar{J}}^{(\vartheta)}
|_{Q^I=\bar{Q}^{\bar{I}}=0} \ne 0$ 
for $J \ne I$ in a separate work~\cite{AKO}.} 
Note that there is no flavor mixing in 
the bulk K\"ahler potential.
Then we obtain 
\begin{eqnarray}
Y_{I\bar{J}}(T,\bar{T},X,\bar{X}) &=& 0 
\qquad (J \ne I). 
\label{eq:nfm} 
\end{eqnarray}
Moreover, when $\Omega^{(0),(\pi)}$ and their derivatives 
are sufficiently small, we have 
\begin{eqnarray}
 & &  \alpha = \alpha_X = \alpha_I = \tilde \alpha_I=
   \beta = \beta_X = \beta_I =\tilde \beta_I =1. 
\nonumber
\end{eqnarray}
The K\"ahler potential is calculated from $\Omega$ as 
\begin{eqnarray}
K &=& 
-3 \ln \left( -\Omega/3 \right) 
\ = \ 
\hat{K}(T,\bar{T},X,\bar{X})
+{\cal O}(Q^2), 
\nonumber
\end{eqnarray}
where 
\begin{eqnarray}
\hat{K}(T,\bar{T},X,\bar{X}) &=& 
\hat{K}_0(T,\bar{T}) 
+Z_{X\bar{X}}(T,\bar{T})|X|^2 
+{\cal O}(|X|^4), 
\label{eq:hsk}
\end{eqnarray}
and 
\begin{eqnarray}
\hat{K}_0(T,\bar{T}) &=& 
-3 \ln \left( -\hat\Omega_0(T,\bar{T})/3 \right), 
\nonumber \\
Z_{X\bar{X}}(T,\bar{T}) &=& 
-3Y_{X\bar{X}}(T,\bar{T}) / \hat\Omega_0(T,\bar{T}). 
\nonumber
\end{eqnarray}

Similarly, we also expand the superpotential $W$ 
and the gauge kinetic function $f_a$ as 
\begin{eqnarray}
W &=& \hat{W}(T,X) 
+\frac{1}{6} \lambda_{IJK}(T,X) Q^IQ^JQ^K 
+{\cal O}(Q^4), 
\nonumber \\
f_a &=& 
f_a^{(0)}|_0+f_a^{(\pi)}|_0+k_aT 
+\left( f_{a,X}^{(0)}|_0+f_{a,X}^{(\pi)}|_0\, 
e^{-(c_X-3k/2)T} \right) X 
+{\cal O}(X^2), 
\label{eq:fa}
\end{eqnarray}
where 
\begin{eqnarray}
\hat{W}(T,X) &=& 
c-\sum_{n=0}^{n_{\rm np}} A_n e^{-a_nT} 
+\left( W_X^{(0)}|_0+W_X^{(\pi)}|_0\, 
e^{-(c_X+3k/2)T} \right) X 
+{\cal O}(X^2), 
\label{eq:hsw} \\
\lambda_{IJK}(T,X) &=& 
W_{IJK}^{(0)}|_0+W_{IJK}^{(\pi)}|_0\, 
e^{-(c_I+c_J+c_K-3k/2)T} 
\nonumber \\ && \qquad 
+\left( W_{IJK,X}^{(0)}|_0+W_{IJK,X}^{(\pi)}|_0\, 
e^{-(c_I+c_J+c_K+c_X-3k)T} \right) X 
+{\cal O}(X^2). 
\label{eq:lijk}
\end{eqnarray}
Note that the constant terms in $W^{(\vartheta)}$ 
are now encoded in the constant $c$ and the zeroth component of $A_n$ as 
\begin{eqnarray}
c &=& W^{(0)}|_0+\sum_{m=1}^{m_{\rm np}} B_m e^{-b_m}, 
\nonumber \\
A_n &=& \{ W^{(\pi)}|_0,A_1,A_2,\ldots,A_{n_{\rm np}} \}, \qquad 
a_n \ = \ \{ 3k,a_1,a_2,\ldots,a_{n_{\rm np}} \}, 
\nonumber
\end{eqnarray}
where $n=0,1,2,\ldots,n_{\rm np}$. 

In the above expressions, we find the K\"ahler/superpotential 
terms for the modulus $T$ and the hidden sector field $X$, 
which carry the informations of dynamical SUSY breaking and 
the uplifting structures. We can also compute soft supersymmetry 
breaking terms from the above expressions, in terms of the 
F-component of the radius modulus $T$ and the hidden sector 
field $X$. With the resulting soft terms, we can analyze the 
nature of sequestering in our 5D model.

\section{Hidden sector potential and soft terms}
\label{sec:soft}
In order to obtain a SUSY breaking Minkowski vacuum with 
(almost) vanishing vacuum energy, we consider the scenario 
of F-term uplifting, which would be realized by the 
scalar potential of the modulus and the hidden sector: 
\begin{eqnarray}
V_F &=& \hat{K}_{i\bar{j}}F^i\bar{F}^{\bar{j}}-3|m_{3/2}|^2, 
\nonumber
\end{eqnarray}
where $i,j,\ldots=(T,X)$, 
\begin{eqnarray}
F^i &=& 
-e^{\hat{K}/2}\hat{K}^{\bar{j}i}
(\bar{\hat{W}}_{\bar{j}}+\hat{K}_{\bar{j}} \bar{\hat{W}}), \qquad 
m_{3/2} \ = \ e^{\hat{K}/2} \hat{W}, 
\nonumber
\end{eqnarray}
and 
$\hat{K}_{i\bar{j}} \hat{K}^{\bar{j}k}=\delta_i^{\ k}$, 
$\hat{K}^{\bar{i}j} \hat{K}_{j\bar{k}}=\delta^{\bar{i}}_{\ \bar{k}}$. 
By substituting $\hat{K}(T,\bar{T},X,\bar{X})$ and $\hat{W}(T,X)$ 
shown in Eqs.~(\ref{eq:hsk}) and (\ref{eq:hsw}), respectively, 
we obtain the modulus and the hidden sector F-term scalar potential 
of our model. 
Note that the above potential is evaluated in the Einstein frame 
where $|C|^2=e^{K/3}M_{Pl}^2$, and we measure all the mass scales 
in the unit $M_{Pl}=1$ in the following. We also restrict ourselves 
to the region $X \ll 1$ where the expansion in powers of $X$ is valid. 

As for the tree-level soft SUSY breaking terms of the visible fields, 
in this paper we focus on\footnote{The so-called $\mu$-term 
and B-term would be discussed in a separate work~\cite{AKO}.} 
the gaugino masses $M_a$, the scalar masses 
$m^2_{I\bar{J}}$ and the A-terms $A_{IJK}$. 
These are defined as 
\begin{eqnarray}
{\cal L}_{\rm soft} &=& 
-m_I^2 |Q^I|^2 
-\frac{1}{2} \left( M_a \lambda^a \lambda^a 
+\frac{1}{6}y_{IJK}A_{IJK}Q^IQ^JQ^K 
+\textrm{h.c.} \right), 
\nonumber
\end{eqnarray}
where 
{\it all the kinetic terms are canonically normalized}, 
and $y_{IJK}=Y_{I\bar{I}}^{-1/2} Y_{J\bar{J}}^{-1/2} 
Y_{K\bar{K}}^{-1/2} \lambda_{IJK}$ is the physical Yukawa coupling. 
Note that there is no flavor mixing in the 
soft scalar masses $m_{I\bar{J}}^2=\delta_{IJ}m_I^2$ 
due to Eq.~(\ref{eq:nfm}). 
These soft terms are generated through the mediation 
by the radius modulus $T$ as well as the direct couplings 
to the SUSY breaking field $X$. Such effects are 
summarized in the following general 
formula~\cite{Kaplunovsky:1993rd,Choi:2005ge}: 
\begin{eqnarray}
M_a &=& F^i \partial_i \ln ({\rm Re}\,f_a), 
\nonumber \\
m_I^2 &=& 
-F^i \bar{F}^{\bar{j}} 
\partial_i \partial_{\bar{j}} \ln Y_{I\bar{I}}, 
\nonumber \\
A_{IJK} &=& F^i \partial_i \ln \left( 
Y_{I\bar{I}}^{-1} Y_{J\bar{J}}^{-1} Y_{K\bar{K}}^{-1} 
\lambda_{IJK} \right), 
\nonumber
\end{eqnarray}
where $i,j,\ldots=(T,X)$. 
Here we assume that the total vacuum energy is vanishing 
at the minimum where the soft terms are evaluated, which 
would be realized by the F-term uplifting. 

Substituting $f_a$, $Y_{I\bar{I}}$ and $\lambda_{IJK}$ 
shown in Eqs.~(\ref{eq:fa}), (\ref{eq:sswf}) and (\ref{eq:lijk}), 
respectively, we find the expressions for the above soft terms as 
\begin{eqnarray}
M_a &=& 
\frac{F^T+(2/k_a)(f_{a,X}^{(0)}
+f_{a,X}^{(\pi)}e^{-(c_X-3k/2)T})\,F^X}{
T+\bar{T}+(2/k_a)(f_a^{(0)}+f_a^{(\pi)})} 
+{\cal O}(X), 
\nonumber \\*[5pt]
m_I^2 &=& 
\beta_I (c_I-k/2)^2 \, 
\frac{|e^{(c_I-k/2)T}|^2}{
(|e^{(c_I-k/2)T}|^2-\beta_I)^2}\, |F^T|^2 
\nonumber \\ && \qquad 
-\frac{\tilde\alpha_I}{3 \alpha_I}\, 
\frac{c_I-k/2}{c_I+c_X-2k}\, 
\frac{|e^{(c_I-k/2)T}|^2-\tilde\beta_I 
|e^{(c_X-3k/2)T}|^2}{
|e^{(c_I-k/2)T}|^2-\beta_I}\, |F^X|^2 
+{\cal O}(X), 
\nonumber \\*[5pt]
A_{IJK} &=& 
-\Bigg\{ \left( \frac{\beta_I(c_I-k/2)}{
|e^{(c_I-k/2)T}|^2-\beta_I} 
+(I \leftrightarrow J) +(I \leftrightarrow K) \right) 
\nonumber \\ && \qquad 
+\frac{c_I+c_J+c_K-3k/2}{
W^{(0)}_{IJK}\, e^{(c_I+c_J+c_K-3k/2)T} 
+W^{(\pi)}_{IJK}} \Bigg\}\, F^T 
\nonumber \\ && \qquad \qquad 
+\frac{W^{(0)}_{IJK,X}
+W^{(\pi)}_{IJK,X}\, e^{-(c_I+c_J+c_K+c_X-3k)T}}{
W^{(0)}_{IJK}
+W^{(\pi)}_{IJK}\, e^{-(c_I+c_J+c_K-3k/2)T}}\, F^X 
+{\cal O}(X), 
\label{eq:genest}
\end{eqnarray}
where we omit the symbol $|_0$. 
In our 5D model, the visible sector fields $Q_I$ and the hidden 
sector field $X$ are quasi-localized with the wavefunction 
$e^{-\hat{c}_Iy}$ and $e^{-\hat{c}_Xy}$, respectively, whose 
effects are encoded in the above expressions as exponential factors. 
In order to suppress the contributions from the direct coupling, 
i.e., $F^X$ in the soft terms, it is favored that $Q^I$ and 
$X$ are localized away from each other. 

Taking into account the warp factor $e^{-\hat{k}y}$ of the 
background geometry, we have basically two choices of kink 
mass parameters for such sequestering of the hidden sector, 
\begin{eqnarray}
&& \left\{ 
\begin{array}{crcllcrcl}
\textrm{(i)} & 
c_I-\frac{k}{2} &\equiv& \tilde{c}_I \ > \ 0 
& (^\forall I), & & 
c_X-\frac{k}{2} &\equiv& -\tilde{c}_X \ < \ 0, \\*[5pt]
\textrm{(ii)} & 
c_I-\frac{k}{2} &\equiv& -\tilde{c}_I \ < \ 0 
&  (^\forall I), & & 
c_X-\frac{k}{2} &\equiv& \tilde{c}_X \ > \ 0. 
\end{array}
\right. 
\label{eq:case1}
\end{eqnarray}
Without loss of generality, we can assume that 
\begin{eqnarray}
k &>& 0. 
\nonumber
\end{eqnarray}
The opposite case $k<0$ is achieved by exchanging 
the quantities originating from two fixed points 
$y=0$ and $y=\pi R$ each other.

\subsection{UV uplifting}
First we consider the case (i) defined in Eq.~(\ref{eq:case1}). 
In this case, the hidden sector field $X$ is localized 
toward the $y=0$ (UV) fixed point. 
The hidden-sector K\"ahler potential (\ref{eq:hsk}) 
and the superpotential (\ref{eq:hsw}) are determined by 
\begin{eqnarray}
\hat{K}_0(T,\bar{T}) &=& 
-3 \ln \left( 
\alpha \, \frac{1-\beta |\epsilon_k(T)|^2}{2k} \right), \qquad 
Z_{X \bar{X}}(T,\bar{T}) \ = \ 
\frac{2k \alpha_X}{\alpha \tilde{c}_X}\, 
\frac{1-\beta_X|\epsilon_X(T)|^2}{1-\beta |\epsilon_k(T)|^2}, 
\nonumber \\
\hat{W}(T,X) &=& 
c-\sum_{n=0}^{n_{\rm np}} A_n e^{-a_nT} 
+\left\{ W^{(0)}_X|_0 +W^{(\pi)}_X|_0 
\epsilon_k^2(T) \epsilon_X(T) \right\} X 
+{\cal O}(X^2). 
\nonumber
\end{eqnarray}
Here and hereafter we use epsilon parameters 
\begin{eqnarray}
\begin{array}{rclcrclcrcl}
\epsilon_I(T) &=& e^{-\tilde{c}_IT}, & & 
\epsilon_X(T) &=& e^{-\tilde{c}_XT}, & & 
\epsilon_k(T) &=& e^{-kT}, 
\end{array}
\nonumber
\end{eqnarray}
whose vacuum values can be exponentially suppressed. 
Especially, $\epsilon_k(\pi R)$ determines the scale at 
infra-red (IR) boundary $y=\pi R$ and also the Kaluza-Klein (KK) 
resonance scale $M_{KK} = {\cal O}(\epsilon_k(\pi R)k)$. 
We compare the above K\"ahler potential and the 
superpotential\footnote{We implicitly assume some heavy modes 
living at the fixed points in our model, which generate a 
SUSY breaking mass $m_X$ for $X$ at the one-loop level as 
in the ISS-KKLT or ISS-racetrack model shown 
in Eq.~(\ref{eq:mx}) in Appendix~\ref{app:ftu}.} 
with those of the ISS-KKLT model (\ref{eq:isskklt}) or the 
ISS-racetrack model (\ref{eq:issrt}) reviewed as basic models 
of F-term uplifting in Appendix~\ref{app:ftu}. 
In the following, we assume an (approximate) $R$-symmetry 
in the hidden sector by assigning the $R$-charge $2$ for $X$. 
Then the quadratic and higher powers of $X$ in the hidden sector 
superpotential are forbidden (or suppressed\footnote{
If there exist such $R$-symmetry breaking terms with higher 
powers of $X$, SUSY vacua would exist in our model. However, 
such SUSY points are far away from the SUSY breaking local 
minimum if the coefficients of the $R$-breaking terms are 
suppressed~\cite{Intriligator:2007cp}.}). This is a 
requirement for a dynamical SUSY breaking~\cite{Nelson:1993nf}. 

For example, for 
\begin{eqnarray}
A_0 &=& W^{(0)}|_0 \ = \ 0, \qquad 
n_{\rm np} \ = \ 1, \qquad 
W^{(\pi)}_X|_0 \ = \ 0, 
\nonumber
\end{eqnarray}
the above superpotential is in the same form as the ISS-KKLT 
model (\ref{eq:isskklt}), with the identification 
\begin{eqnarray}
A_1 &=& A, \qquad 
a_1 \ = \ a, \qquad 
W^{(0)}_X|_0 \ = \ \mu^2. 
\nonumber
\end{eqnarray}
Only the difference is that the above modulus K\"ahler 
potential $\hat{K}_0(T,\bar{T})$ carries $\epsilon_k(T)$ due 
to the warped background geometry. In the limit $k \to 0$, 
this is reduced to 
$\hat{K}_0(T,\bar{T})=-3\ln(T+\bar{T})
+{\cal O}(\Omega^{(0)}|_0,\Omega^{(\pi)}|_0)$, that is, 
to the effective modulus K\"ahler potential of the KKLT model 
(\ref{eq:isskklt}) with some corrections from boundary constants. 
However, even with a finite value of $k$, as long as 
the following relation,
\begin{eqnarray}
1-a(\partial_T \hat{K}_0)^{-1}|_{T=T_0} 
&=& 1-(3k)^{-1}a(1-|\epsilon_k(T_0)|^{-2}) 
\ \ll \ A_1c^{-1}, 
\nonumber
\end{eqnarray}
is satisfied (see Eq.(\ref{eq:dtw0})) and also 
the $T$-$X$ mixing is small as in Eq.~(\ref{eq:mixcd}), 
the reference point, 
\begin{eqnarray}
T_0 &\simeq& a^{-1} \ln(Ac^{-1}), \qquad 
X_0 \ = \ {\cal O}((\mu/m_X)^2 c), 
\nonumber
\end{eqnarray} 
satisfying 
$\hat{W}_T+\hat{K}_T \hat{W} = 0$, 
$V_X=0$ ($\hat{W}_X+\hat{K}_X \hat{W} \ne 0$), 
almost represents the SUSY breaking minimum 
(see Eqs.~(\ref{eq:refpt}) and (\ref{eq:refcd})) 
up to 
certain small deviations $\delta T$ and $\delta X$ 
as in the ISS-KKLT model shown in Appendix~\ref{app:ftu}. 
Therefore, if the parameters satisfy a weak warping condition 
\begin{eqnarray}
c &\ll& |\epsilon_k(T_0)|^2, \qquad 
\textrm{i.e.,} \qquad 2kT_0 \ \ll \ \ln c^{-1}, 
\label{eq:ww}
\end{eqnarray}
as well as the low energy SUSY conditions 
\begin{eqnarray}
\ln c^{-1} &=& {\cal O}(4 \pi^2), \qquad 
\mu^2 \ = \ {\cal O}(\tilde{c}), 
\nonumber
\end{eqnarray}
(see Eqs.~(\ref{eq:clf}) and (\ref{eq:mu0c}), respectively), 
we can realize a kind of the ISS-KKLT model effectively in our 5D framework. 

The vacuum energy at the SUSY breaking minimum is vanishing 
if we tune the corresponding parameters as in Eq.~(\ref{eq:vve}), i.e., 
\begin{eqnarray}
\tilde{c} &\simeq& 
\mu^2/\sqrt{3 Z_{X\bar{X}}}|_{T=T_0}
+{\cal O}(\mu^2), 
\nonumber
\end{eqnarray}
where the SUSY AdS vacuum in the modulus $T$ sector alone 
is uplifted to a SUSY breaking Minkowski minimum by a dynamically 
generated F-term in the hidden sector $X$. 
With negligible K\"ahler mixing 
$\hat{K}_{X\bar{T}},\hat{K}_{XT} = {\cal O}(X) \ll 1$, 
the modulus mass $m_T$ and the SUSY breaking order parameters 
$F^T$, $F^X$, $F^C$ are expressed (see Eq.~(\ref{eq:genft})) as 
\begin{eqnarray}
m_T &\simeq& 
-e^{\hat{K}/2} \hat{K}^{T \bar{T}} \hat{W}_{TT}|_{T=T_0,X=X_0}, 
\nonumber \\
F^T &\simeq& \left. 
-\frac{\sqrt{3} \hat{K}_T}{\hat{K}_{T \bar{T}}} \left( 
\sqrt{3}+\frac{\sqrt{\hat{K}^{X \bar{X}}} \hat{W}_{TX}}{
\hat{K}_T \hat{W}} \right) \frac{|m_{3/2}|^2}{m_T} 
\right|_{T=T_0,X=X_0}, 
\nonumber \\
F^X &\simeq& 
\left. -\sqrt{\frac{3}{\hat{K}_{X \bar{X}}}}\,
m_{3/2} \right|_{T=T_0,X=X_0}, 
\qquad 
F^C \ \simeq \ \left. C m_{3/2} \right|_{T=T_0,X=X_0}. 
\nonumber
\end{eqnarray}
Because we are considering a weak warping (\ref{eq:ww}), 
$m_T$, $F^T$, $F^C$  and $F^X$ can be also estimated by 
those in the ISS-KKLT model, 
(\ref{eq:kkltmt}), (\ref{eq:kkltft}) and (\ref{eq:fx}), 
respectively. 
That is, the modulus $T$ is heavier than the gravitino 
by a loop factor, and the tree-level modulus mediation 
is comparable to the anomaly mediation, 
\begin{eqnarray}
m_T &=& {\cal O}(4 \pi^2 m_{3/2}), \quad 
\frac{F^T}{T+\bar{T}} \ = \ {\cal O}(m_{3/2}/4 \pi^2), \quad 
F^X \ \sim \ \frac{F^C}{C} \ = \ {\cal O}(m_{3/2}). 
\label{eq:mirage}
\end{eqnarray}

On the other hand, for 
\begin{eqnarray}
c &=& 0, \qquad 
A_0 \ = \ W^{(0)}|_0=0, \qquad 
n_{\rm np} \ = \ 2, \qquad 
W^{(0)}_X|_0=0, 
\nonumber
\end{eqnarray}
the above superpotential has the same form as one in the 
ISS-racetrack model (\ref{eq:issrt}), if we tune the parameters as, 
e.g., $kT_0 \ll \tilde{c}_XT_0 = {\cal O}(4 \pi^2)$. 
However, in this case, a heavier modulus mass 
can be obtained\footnote{This enhancement of 
modulus mass would play a role to avoid the so-called moduli-induced 
gravitino/neutralino problem~\cite{Nakamura:2006uc} 
in the KKLT-type scenario. 
Note that the modulus mass is already enhanced by a 
loop factor in the ISS-racetrack model (\ref{eq:rtmt}) 
compared with the original KKLT model~\cite{Abe:2007yb}.} 
without affecting the size of $F^T$ and $F^C$, that is,  
\begin{eqnarray}
m_T &=& {\cal O}(\epsilon_k^{-2}(4 \pi^2)^2 m_{3/2}), \quad 
\frac{F^T}{T+\bar{T}} \ = \ {\cal O}(m_{3/2}/4 \pi^2), \quad 
F^X \ \sim \ \frac{F^C}{C} \ = \ {\cal O}(m_{3/2}). 
\nonumber
\end{eqnarray}
Note that the modulus mass $m_T$ is heavier by 
a factor $\epsilon_k^{-2}$  than 
the one in the ISS-racetrack model (\ref{eq:rtmt}).

So far, we have confirmed that our 5D model can realize the 
ISS-KKLT-type moduli stabilization, SUSY breaking and F-term 
uplifting, under the assumption of weak warping (\ref{eq:ww}). 
It might be possible to construct a different class of F-term 
uplifting model with a strong warping, e.g., 
$k = {\cal O}(4 \pi^2)$ and then $\epsilon_k = {\cal O}(c)$. 
In this case, we have to be careful about the fact that the KK 
scale is quite low and the effects of non-zero modes can be 
enhanced at low energy. This case is beyond the scope of 
this paper, and we would study it elsewhere. 
(For the case with $k=a/3 = {\cal O}(4 \pi^2)$, 
see Ref.~\cite{Abe:2007zv}.) 

Next, we study the nature of sequestering in the case (i), 
where $Q^I$ are localized toward the $y=\pi R$ (IR) fixed point. 
The tree-level soft terms for visible fields are found as 
\begin{eqnarray}
M_a &=& 
\frac{F^T+k_a^{-1} \left( f_{a,X}^{(0)} 
+\epsilon_k^{-1} \epsilon_X 
f_{a,X}^{(\pi)} \right) F^X}{
T+\bar{T}+2k_a^{-1} (f_a^{(0)}+f_a^{(\pi)})} 
+{\cal O}(X), 
\nonumber \\
m_I^2 &=& 
\frac{\beta_I \tilde{c}_I^2}{\beta_I-\epsilon_I^2}\, 
\left( \frac{\epsilon_I^2}{\beta_I-\epsilon_I^2}\, 
|F^T|^2 
-\frac{\tilde\alpha_I}{3 \alpha_I \beta_I}\, 
\frac{\epsilon_I^2-\tilde\beta_I 
\epsilon_k^{-1} \epsilon_X}{
\tilde{c}_I (\tilde{c}_X-\tilde{c}_I-k)}\, 
|F^X|^2 \right) +{\cal O}(X), 
\nonumber \\
A_{IJK} &=& 
-\left\{ \left( \frac{\beta_I \tilde{c}_I}{
\beta_I-\epsilon_I^2} 
+(I \leftrightarrow J)+(I \leftrightarrow K) \right) 
-\frac{(\tilde{c}_I+\tilde{c}_J+\tilde{c}_K) 
W^{(\pi)}_{IJK}}{
W^{(0)}_{IJK} \epsilon_I \epsilon_J \epsilon_K 
+W^{(\pi)}_{IJK}} 
\right\} F^T 
\nonumber \\ && \qquad 
+\frac{W^{(0)}_{IJK,X} 
\epsilon_I \epsilon_J \epsilon_K 
+W^{(\pi)}_{IJK,X} \epsilon_k^{-1} \epsilon_X}{
W^{(0)}_{IJK} \epsilon_I \epsilon_J \epsilon_K 
+W^{(\pi)}_{IJK}}\, F^X 
+{\cal O}(X), 
\label{eq:uvust}
\end{eqnarray}
where we omit the symbol $|_0$. 

We assume that the order parameters $F^T$, $F^C$ and $F^X$ 
are given by (\ref{eq:mirage}). 
For $f_{a,X}^{(0)} = {\cal O}(1)$, we find 
$M_a = {\cal O}(m_{3/2})$. 
If we assume that the hidden sector $R$-symmetry is preserved 
also in the visible sector gauge kinetic functions, 
i.e., $f_{a,X}^{(0)}=f_{a,X}^{(\pi)}=0$, 
the $F^X$ can not contribute to the gaugino mass. 
In this case, the gaugino mass is a mirage-type, where the 
tree-level modulus mediation and the anomaly mediation 
are comparable to each other. 

The tree-level contributions to scalar masses $m_I^2$ 
are suppressed compared with the anomaly mediation 
if $\epsilon_I^2$, $\epsilon_k^{-1}\epsilon_X \ll 1/(4 \pi^2)$. 
In this case, the sequestering is maximal.
The limit $\tilde{c}_I \to \infty$, that is, 
$\epsilon_I \to 0$, corresponds to the complete localization 
of $Q^I$  at the fixed point $y=\pi R$. 
Then, $Q_I$ is a chiral multiplet living only at 
the fixed point $y=\pi R$ and its scalar mass $m_I$ 
has no contribution due to $F^T$.
Similarly, the limit $\tilde{c}_X \to \infty$ makes
$X$ live only at the fixed point $y=0$, 
and the scalar mass $m_I$ of $Q_I$ has no 
contribution due to $F^X$.
At any rate, the modulus mediation is always 
subdominant compared with the direct mediation unless all the 
epsilon parameters are of ${\cal O}(1)$. 
Note that the modulus 
mediation and the direct mediation typically give a positive 
and a negative contribution to scalar masses squared, respectively.


The contact term between $X$ and $Q^I$  would be 
induced by loop effects through the gravitational interaction 
even if $X$ and $Q^I$ are completely localized at 
opposite fixed points~\cite{Gherghetta:2001sa}.
Such loop effects would lead to corrections to scalar masses 
squared $\Delta m_I^2$, which are proportional to 
$|F^X|^2$.
However, such corrections are suppressed by 
the one-loop factor and the warp factor $\epsilon_k^2$.
(See also Ref.~\cite{Choi:1997de}.)\footnote{
Other sources of contact terms would be bulk vector 
multiplets~\cite{Brummer:2006dg,Choi:2006bh}, 
which are (assumed to be) absent in our 5D model.}
Thus, such corrections are negligible compared 
with the anomaly mediation in a weakly warped case, e.g. 
$\epsilon_k  \sim M_{GUT}/M_{Pl}$ and even in the case with 
$\epsilon_k  \sim 1/(4 \pi^2)$.

For $\epsilon_I^2$ $(^\forall I)$, 
$\epsilon_I \epsilon_J \epsilon_K$, 
$\epsilon_k^{-1} \epsilon_X \ll 1/(4 \pi^2)$, 
the tree-level contributions to the A-term are suppressed 
compared with the anomaly mediation if $W^{(\pi)}_{IJK}|0 \ne 0$, 
and the maximal sequestering is achieved. 
However, if $W^{(\pi)}_{IJK}|_0=0$, the A-term becomes 
$A_{IJK} \simeq -(\tilde{c}_I+\tilde{c}_J+\tilde{c}_K)F^T$ 
\cite{Choi:2003di,Abe:2004tq}
which can be a mirage-type for 
$\tilde{c}_I$, $\tilde{c}_J$, $\tilde{c}_K = {\cal O}(1)$.

\subsection{IR uplifting}
Next we consider the case (ii) defined in Eq.~(\ref{eq:case1}). 
In this case, the hidden sector field $X$ is localized 
toward the $y=\pi R$ (IR) fixed point. 
For $k=0$, i.e., a flat extra dimension, the case (ii) is 
physically equivalent to the case (i) under the exchange 
of two fixed points. The difference between the case (i) 
and the case (ii) is enhanced for large $k$. Thus, we only consider 
such a case, 
\begin{eqnarray}
\epsilon_k &\lesssim& \epsilon_X,\ \epsilon_I, \qquad 
(\textrm{i.e. } k \gtrsim \tilde{c}_X, \tilde{c}_I), 
\label{eq:sw}
\end{eqnarray}
for $^\forall I$ in the following. 
In this case, the hidden sector (and the modulus) 
K\"ahler and the superpotential are obtained as 
\begin{eqnarray}
\hat{K}(T,\bar{T},\tilde{X},\bar{\tilde{X}}) &=& 
-3 \ln \left( 
\alpha \, \frac{1-\beta |\epsilon_k(T)|^2}{2k} \right) 
+\frac{2k \alpha_X}{\alpha \tilde{c}_X}\, 
\frac{\beta_X-|\epsilon_X(T)|^2}{1-\beta |\epsilon_k(T)|^2}\, 
|\tilde{X}|^2 +{\cal O}(|\tilde{X}|^4), 
\nonumber \\
\hat{W}(T,\tilde{X}) &=& 
c-\sum_{n=0}^{n_{\rm np}} A_n e^{-a_nT} 
+\left\{ W^{(0)}_X|_0 \epsilon_X(T) 
+W^{(\pi)}_X|_0 \epsilon_k^2(T) \right\} \tilde{X} 
+{\cal O}(\tilde{X}^2), 
\nonumber
\end{eqnarray}
where we redefined $X$ as 
\begin{eqnarray}
\tilde{X} &=& \epsilon_X^{-1}(T)X. 
\nonumber
\end{eqnarray}

The superpotential can be the ISS-KKLT type (\ref{eq:isskklt}) 
or the ISS-racetrack type (\ref{eq:issrt}), for the case of 
$W^{(0)}_X|_0=0$ or $W^{(\pi)}_X|_0=0$ with the identification 
$\epsilon_X(T)W^{(0)}_X|_0=\mu^2(T)$ or 
$\epsilon_k^2(T)W^{(\pi)}_X|_0=\mu^2(T)$, respectively, 
where a sizable $T$-$X$ mixing exists for 
$k \gtrsim \tilde{c}_X \gtrsim 4 \pi^2$. 
As shown previously, the effect of warping in the 
K\"ahler potential is not relevant to the analysis 
in Appendix~\ref{app:ftu}, only when a weak warping 
condition (\ref{eq:ww}) is satisfied, and  
this restricts $\epsilon_X$ and $\epsilon_I$ as 
$c \ll \epsilon_X$, $\epsilon_I$ due to Eq.~(\ref{eq:sw}). 
Then the corresponding condition to Eq.~(\ref{eq:mu0c}) 
can not be satisfied, that is, the analysis in 
Appendix~\ref{app:ftu} is not valid in this case, and 
the structure of F-term uplifting can be different from 
the conventional one due to Eq.~(\ref{eq:sw}). 
We would study also this case elsewhere as well as the 
above mentioned strong warping case. 
At any rate, for the weak warping (\ref{eq:ww}) {\it without} 
the condition (\ref{eq:sw}), the physics should be almost 
the same as the case (i) as mentioned above. Then we would 
realize the ISS-KKLT model and the ISS-racetrack model 
effectively, also in the case (ii). 

The tree-level soft terms for visible fields in the case (ii) 
are derived as 
\begin{eqnarray}
M_a &=& 
\frac{F^T+2k_a^{-1} \left( 
\epsilon_X f_{a,X}^{(0)} +\epsilon_k^{-1} 
f_{a,X}^{(\pi)} \right) F^{\tilde{X}}}{
T+\bar{T}+2k_a^{-1} (f_a^{(0)}+f_a^{(\pi)})} 
+{\cal O}(\tilde{X}), 
\nonumber \\
m_I^2 &=& 
\frac{\beta_I \tilde{c}_I^2}{1-\beta_I \epsilon_I^2}\, 
\left( \frac{\epsilon_I^2}{1-\beta_I \epsilon_I^2}\, 
|F^T|^2 
-\frac{\tilde\alpha_I}{3\alpha_I \beta_I}\, 
\frac{\epsilon_X^2-\tilde\beta_I 
\epsilon_k^{-1} \epsilon_I}{
\tilde{c}_I (\tilde{c}_I-\tilde{c}_X-k)}\, 
|F^{\tilde{X}}|^2 \right) 
+{\cal O}(\tilde{X}), 
\nonumber \\
A_{IJK} &=& 
-\left\{ \left( \frac{\beta_I \tilde{c}_I \epsilon_I^2}{
1-\beta_I \epsilon_I^2} 
+(I \leftrightarrow J)+(I \leftrightarrow K) \right) 
+\frac{(\tilde{c}_I+\tilde{c}_J+\tilde{c}_K) 
W^{(\pi)}_{IJK} \epsilon_I \epsilon_J \epsilon_K}{
W^{(0)}_{IJK} 
+W^{(\pi)}_{IJK} \epsilon_I \epsilon_J \epsilon_K} 
\right\} F^T 
\nonumber \\ && \qquad 
+\frac{W^{(0)}_{IJK,X} \epsilon_X 
+W^{(\pi)}_{IJK,X} \epsilon_k^{-1} 
\epsilon_I \epsilon_J \epsilon_K}{
W^{(0)}_{IJK}+W^{(\pi)}_{IJK} 
\epsilon_I \epsilon_J \epsilon_K}\, F^{\tilde{X}} 
+{\cal O}(\tilde{X}), 
\nonumber
\end{eqnarray}
where we again omit the symbol $|_0$. 
The difference from Eq.~(\ref{eq:uvust}) is just the position 
of $\epsilon_k^{-1}$ factors aside from the exchange of 
$y=0$ and $y=\pi R$ fixed points with each other. 

If we consider the case satisfying Eq.~(\ref{eq:sw}) 
(although above we have not showed the corresponding 
vacuum in the hidden sector), the contribution from 
the direct coupling to the gaugino mass $M_a$ is 
enhanced by $\epsilon_k^{-1}$ if $f_{a,X}^{(\pi)} \ne 0$, 
the scalar mass $m_I$ is of ${\cal O}(F^{\tilde{X}})$ or larger, 
but the A-term can be suppressed if $W^{(0)}_{IJK} \ne 0$.

\subsection{Patterns of soft terms}

Here, let us summarize 
resultant patterns of soft SUSY breaking terms 
in our model.
First recall the sizes of $F^T$, $F^X$ and $F^C$, 
(\ref{eq:mirage}), that is, $F^X$ is larger than $F^T$ 
by a factor of ${\cal O}(4 \pi^2)$ and 
the modulus mediation is comparable to the 
anomaly mediation.
When $f_{a,X}^{(0)}$, 
$f_{a,X}^{(\pi)}e^{-(c_X-3k/2)T}= {\cal O}(1)$, and 
$e^{(c_I-k/2)T}$ or $e^{(c_X-3k/2)T}$ is not suppressed 
in Eq.~(\ref{eq:genest}), the F-term of $X$, $F^X$, is 
dominant in all of soft terms, which are obtained as 
\begin{eqnarray}
 & & M_a = {\cal O}(m_{3/2}), \qquad 
m^2_I = {\cal O}(m_{3/2}^2), \qquad 
A_{IJK} = {\cal O}(m_{3/2}),
\nonumber
\end{eqnarray}
that is, the visible sector is not sequestered from the 
dominant SUSY breaking source.
Their explicit ratios to $m_{3/2}$ are model-dependent like 
generic spectrum due to the gravity mediation.

When $f_{a,X}^{(0),(\pi)}$ are sufficiently suppressed, 
the size of gaugino masses is estimated as 
$M_a={\cal O}(m_{3/2}/(4\pi^2))$.
On the other hand, unless $e^{(c_I-k/2)T}$ or $e^{(c_X-3k/2)T}$ 
is not suppressed in Eq.~(\ref{eq:genest}),
the size of scalar masses is estimated as 
$m^2_I = {\cal O}(m_{3/2}^2)$.
However, those are tachyonic in a natural parameter 
region like $\alpha_I \sim \tilde \alpha_I \sim \beta_I \sim 
\tilde \beta_I \sim 1$.
Thus, the contribution of $F^X$ to scalar masses must 
be sequestered except in a certain model like a 
negative value of $\tilde \alpha_I/\alpha_I$ and/or 
$\tilde \beta_I/\beta_I$.
Note that when we suppress the contribution of $F^X$ to scalar masses 
by requiring $e^{(c_I-k/2)T} \ll 1$ as well as 
$e^{(c_X-3k/2)T} \ll 1$, the contribution from $F^T$ is 
always suppressed.\footnote{
This is because our model has only a single extra dimension.
If we would extend our scenario to models with more than 
one extra dimensions, we could obtain soft scalar masses, where 
the contribution from the dominant SUSY breaking $F^X$ is 
sequestered but some moduli F-terms have 
significant contributions.}
In this case, the modulus mediation contribution to the 
gaugino masses is obtained as 
\begin{eqnarray}
M_a &=& 
\frac{F^T}{
T+\bar{T}+(2/k_a)(f_a^{(0)}+f_a^{(\pi)})} = {\cal O}(m_{3/2}/(4 \pi^2)),
\nonumber 
\end{eqnarray}
and the $F^X$ contributions to the scalar masses in the case (i), 
i.e. the UV uplifting (\ref{eq:uvust}), are obtained as 
\begin{eqnarray}
m_I^2 &=&  -\frac{\tilde{c}_I\tilde\alpha_I}{3 \alpha_I \beta_I}\, 
\frac{\epsilon_I^2-\tilde\beta_I 
\epsilon_k^{-1} \epsilon_X} {(\tilde{c}_X-\tilde{c}_I-k)}\, |F^X|^2   .
\nonumber 
\end{eqnarray}
When $\alpha_I = \tilde\alpha_I = \beta_I  = \tilde\beta_I =1$, 
this reduces to 
\begin{eqnarray}
m_I^2 &=&  -\frac{\tilde{c}_I}{3 }\, 
\frac{\epsilon_I^2- 
\epsilon_k^{-1} \epsilon_X} {(\tilde{c}_X-\tilde{c}_I-k)}\, |F^X|^2   .
\nonumber 
\end{eqnarray}
This scalar mass must be suppressed as $|m^2_I| \leq {\cal}O(M^2_a)$ 
as the above reason.
Otherwise, scalar masses become tachyonic even at a low-energy scale 
when we include radiative corrections due to gaugino masses.
Then, the anomaly mediation is comparable, that is, 
the mirage mediation, and magnitudes of soft masses are estimated as 
\begin{eqnarray}
 & & M_a = {\cal O}(m_{3/2}/(4 \pi^2)), \qquad 
m^2_I = {\cal O}(m_{3/2}^2/(4 \pi^2)^2).
\nonumber
\end{eqnarray}

\section{Summary and discussions}
\label{sec:summary}
We studied concretely several issues, 
moduli stabilization, SUSY breaking, F-term uplifting 
and sequestering, altogether in a simple supergravity model 
with a single extra dimension. These issues are realized in a 
fully dynamical way by the use of wavefunction localization 
in extra dimension, allowing explicit calculations. 
Especially, we found that the sequestering in the soft scalar 
mass and the A-term can be achieved within the framework of F-term 
uplifting in our 5D model. Because the radius modulus is stabilized 
by a KKLT-type potential, the gaugino mass is a mirage-type if the 
visible sector gauge kinetic function preserves the hidden sector 
$R$-symmetry which is responsible for the dynamical SUSY breaking, 
and then the tree-level modulus mediation is comparable to the 
anomaly mediation. It is notable that the TeV scale mirage 
mediation~\cite{Choi:2005hd} can solve the so called little 
hierarchy problem~\cite{Barbieri:1987fn} within the MSSM, due to 
the gluino and wino mass unification at the TeV scale~\cite{Abe:2007kf}. 
Note also that our 5D model might have a corresponding 
conformal field theory (CFT) description\footnote{See Ref.~\cite{AKO2} 
for a realization of ISS sector in this direction.} 
due to the AdS/CFT correspondence~\cite{Maldacena:1997re}. 

We have only considered typical cases (\ref{eq:case1}) from 
the viewpoint of sequestering, i.e., all the generations of 
quarks and leptons localize toward the common fixed point. 
However, as mentioned in the introduction, it is known that 
a certain class of realistic Yukawa matrices for quarks and 
leptons can be obtained by the wavefunction localization 
in 5D supergravity theory. In this case, either the light generation 
or the heavy generation would be forced to localize toward 
the same fixed points as the SUSY breaking field $X$. 
For such a generation, sequestering can not occur, and the 
squark/slepton might receive a soft scalar mass with a 
large magnitude from the direct coupling. 
We would study more detailed flavor structure of our model with 
such realistic Yukawa couplings in a separate work~\cite{AKO}. 
(For the case of radion domination or Scherk-Schwarz (SS) SUSY 
breaking~\cite{Scherk:1978ta}\footnote{
Note that the SS twist is prohibited in our 5D supergravity 
model~\cite{Hall:2003yc}. 
We can introduce such a twist in a different type of 
orbifold supergravity~\cite{Altendorfer:2000rr,Abe:2007dh}.}, 
see Refs.~\cite{Choi:2003di,Abe:2004tq} and 
references therein.) 

We also restricted to the case with a single modulus. 
If we consider more $Z_2$-odd vector multiplets in 5D, 
we have multiple moduli which can cause moduli mixing 
in the gauge kinetic function, and then in the 
nonperturbative superpotential. Such mixing effects 
could play important roles in the moduli stabilization~\cite{
Abe:2005rx,Choi:2006bh} after integrating out 
heavy moduli~\cite{Choi:2004sx,Abe:2006xi}. 

Our model is not directly related to a certain string model 
known until now. However, the result of this paper would help 
us to understand some basic features of the moduli stabilization, 
the SUSY breaking, the realization of Minkowski (de Sitter) vacuum 
and the sequestering in higher-dimensional supergravity models and 
superstring models.

\subsection*{Acknowledgement}
The authors would like to thank Kiwoon~Choi for useful discussions.
H.~A.\/,  T.~H.\/ and T.~K.\/ are supported in part by the
Grand-in-Aid for Scientific Research \#182496, \#194494 
and  \#17540251, respectively.
T.~K.\/ is also supported in part by 
the Grant-in-Aid for
the 21st Century COE ``The Center for Diversity and
Universality in Physics'' from the Ministry of Education, Culture,
Sports, Science and Technology of Japan.

\appendix

\section{Basic structure of F-term uplifting}
\label{app:ftu}
In this appendix, we review the scenario of 
F-term uplifting~\cite{Dudas:2006gr,Abe:2006xp,Abe:2007yb} 
based on the KKLT-type SUSY AdS vacua.

\subsection{KKLT model}
First we briefly review the original KKLT 
model~\cite{Kachru:2003aw,Choi:2004sx,Choi:2005ge}. 
The F-term potential of the 4D N=1 supergravity theory is given by 
\begin{eqnarray}
V_F &=& K_{I\bar{J}}F^I\bar{F}^{\bar{J}}-3|m_{3/2}|^2, 
\nonumber
\end{eqnarray}
where $F^I$ is the F-component of the $I$-th chiral multiplet 
and $m_{3/2}$ is the gravitino mass given respectively as 
$F^I=-e^{K/2}K^{I\bar{J}}
(\bar{W}_{\bar{J}}-K_{\bar{J}} \bar{W})$ and 
$m_{3/2}=e^{K/2}W$. 

Before introducing an uplifting sector, the KKLT model 
assumes the following K\"ahler potential and superpotential, 
\begin{eqnarray}
K &=& -3 \ln(T+\bar{T}), \qquad 
W \ = \ c-Ae^{-aT}, 
\label{eq:kklt}
\end{eqnarray}
in the 4D effective supergravity theory, 
where $T$ is a light modulus, 
the first constant term, $c$, originates from a flux, 
the second $T$-dependent term comes from a nonperturbative 
effect such as a gaugino condensation and then 
$a = {\cal O}(4 \pi^2)$, $A = {\cal O}(1)$. 
We measure all the mass scales in the unit $M_{Pl}=1$. 
In order to realize low scale SUSY breaking (TeV scale 
gravitino mass), we consider a tiny flux constant, 
\begin{eqnarray}
\ln c^{-1} &=& {\cal O}(4 \pi^2). 
\label{eq:clf}
\end{eqnarray}
Then, the SUSY stationary condition of the scalar potential, 
$F^T=0$, is satisfied by 
\begin{eqnarray}
aT_0 &=& 
\ln(Ac^{-1})+\ln(1-aK_T^{-1}|_{T=T_0}) 
\ \simeq \ \ln(Ac^{-1}) 
\ = \ {\cal O}(4 \pi^2). 
\label{eq:dtw0}
\end{eqnarray}
This stationary solution corresponds to a SUSY AdS vacuum 
of the scalar potential with a negative vacuum energy 
$V_{SUSY}=\left. -3|m_{3/2}|^2 \right|_{T=T_0} 
= {\cal O}(c^2)$. 

In the original KKLT model, this SUSY AdS minimum is uplifted 
to a Minkowski minimum by introducing an anti D3-brane. 
The anti-brane breaks $N=1$ SUSY explicitly in the 4D effective 
supergravity, and generates an uplifting potential energy 
\begin{eqnarray}
U &=& \int d^4\theta\,|C|^4 
\xi \theta^2 \bar\theta^2
\ = \ \xi e^{2K/3}, 
\label{eq:ad3u}
\end{eqnarray}
where $\xi$ is a constant. 
The total scalar potential is now given by $V=V_F+U$ and 
the previous minimum is shifted as $T=T_0+\delta T$. 
We tune the constant 
$\xi= \left. 3e^{-2K/3}|m_{3/2}|^2 \right|_{T=T_0}$ so that 
$V=0$ at the leading order in a $\delta T/T_0$ expansion. 
Then, we find the shift of $T$ at this Minkowski minimum, 
$\delta T/T_0 \sim 1/(aT_0)^2 = {\cal O}(1/(4 \pi^2)^2)$, 
and the modulus mass, 
\begin{eqnarray}
m_T &\simeq& aT_0\,m_{3/2} 
\ = \ {\cal O}(4 \pi^2 m_{3/2}). 
\label{eq:kkltmt}
\end{eqnarray}
The SUSY breaking order parameters are estimated as 
\begin{eqnarray}
\frac{F^T}{T+\bar{T}} 
&\simeq& \frac{m_{3/2}}{aT_0} 
\ = \ {\cal O}(m_{3/2}/4 \pi^2),  \qquad 
\frac{F^C}{C} \ \simeq \ m_{3/2} \ = \ {\cal O}(c). 
\label{eq:kkltft}
\end{eqnarray}
Here we find that the tree level modulus mediation 
is comparable to the one-loop anomaly mediation, 
$F^T \sim F^C/4 \pi^2$, that is, the so-called 
mirage mediation~\cite{Choi:2005uz}. 

If we consider a racetrack model, 
\begin{eqnarray}
W &=& Ce^{-cT}-Ae^{-aT}, 
\label{eq:rt}
\end{eqnarray}
instead of the KKLT superpotential (\ref{eq:kklt}), the 
modulus mass $m^T$ ($F^T$) is more enhanced (suppressed) as 
\begin{eqnarray}
m_T &\simeq& (aT_0)(cT_0)\,m_{3/2} 
\ = \ {\cal O}((4 \pi^2)^2 m_{3/2}), 
\label{eq:rtmt} \\
\frac{F^T}{T+\bar{T}} 
&\simeq& \frac{m_{3/2}}{(aT_0)(cT_0)} 
\ = \ {\cal O}(m_{3/2}/(4 \pi^2)^2). 
\label{eq:rtft}
\end{eqnarray}
Thus the modulus mediation is negligible compared with the 
anomaly mediation in this case.

\subsection{ISS-KKLT model}
In the original KKLT model, the uplifting potential 
(\ref{eq:ad3u}) is a kind of an explicit SUSY breaking term 
in the low energy effective theory. Instead, we can consider 
the case that the uplifting potential is supplied by the 
$F$-term of a dynamical SUSY breaking sector $X$ which is 
included in the F-term potential $V_F$ itself. 
If $X$ is anyhow sequestered from $T$, the picture that 
the AdS SUSY vacua existing in the $T$ sector alone is 
uplifted by $F^X$ generated by the $X$ sector, would be valid. 
Then, we assume the following K\"ahler and superpotential, 
\begin{eqnarray}
K &=& -3 \ln(T+\bar{T})+Z_{X\bar{X}}(T,\bar{T})|X|^2, \qquad 
W \ = \ c-Ae^{-aT}+\mu^2(T) X. 
\label{eq:isskklt}
\end{eqnarray}
The tadpole of $X$ would appear as a low energy effective 
superpotential term in the dynamical SUSY breaking sector, 
such as the O'Raifeartaigh model~\cite{O'Raifeartaigh:1975pr} 
and the Intriligator-Seiberg-Shih (ISS) model~\cite{Intriligator:2006dd}, 
after integrating out heavy modes, and we call 
the model (\ref{eq:isskklt}) the ISS-KKLT model. 
The effect of the heavy modes appears at low energy 
as a one-loop correction to the above K\"ahler potential, 
\begin{eqnarray}
\Delta K &=& -\Lambda^{-2} Z^{(1)}(T,\bar{T})|X|^4, 
\nonumber
\end{eqnarray}
and then the correction to the scalar potential in this case 
is expressed as a SUSY breaking mass term of $X$, 
\begin{eqnarray}
\Delta V_F &=& m_X^2|X|^2 +{\cal O}(|X|^4), \qquad 
m_X^2 \ = \  e^K(4 \mu^4 Z^{(1)})/(Z_{X \bar{X}} \Lambda)^2, 
\label{eq:mx}
\end{eqnarray}
where $\Lambda$ is the mass scale of the heavy modes.\footnote{
In Ref~\cite{Abe:2007yb}, the $Z_{X\bar{X}}$ 
($Z$ in the notation of Ref.~\cite{Abe:2007yb}) 
dependence of $m_X^2$ should be replaced by 
$Z_{X\bar{X}}^{-2}$, which is a typographical error.} 

If the $T$-$X$ mixing is small, 
\begin{eqnarray}
|K_{T\bar{X}}| 
&\ll& K_{T\bar{T}},\ K_{X\bar{X}} \ = \ {\cal O}(1), \qquad 
|K_{TX}| \ \ll \ |K_{TT}|, \qquad 
|W_{TX}| \ \ll \ |W_{TT}|, 
\label{eq:mixcd}
\end{eqnarray}
the solution 
\begin{eqnarray}
T_0 &\simeq& a^{-1}\ln(Ac^{-1}), \qquad 
X_0 \ \simeq \ 2 (\mu_0/m_X)^2 \tilde{c}, 
\label{eq:refpt}
\end{eqnarray}
satisfying 
\begin{eqnarray}
W_T+K_TW &=& 0, \qquad V_X \ =  \ 0 \quad (W_X+K_X W \ne 0), 
\label{eq:refcd}
\end{eqnarray}
would be a good {\it reference point} of 
the SUSY breaking minimum, where 
$Z_0=Z_{X \bar{X}}(T_0,\bar{T}_0)$, $\mu_0=\mu(T_0)$ 
and $\tilde{c}=c-Ae^{-aT_0}$. Here we have assumed 
\begin{eqnarray}
m_X^2 &=& {\cal O}(\mu_0^2/(4 \pi^2)), \qquad 
\mu_0^2 \ = \ {\cal O}(\tilde{c}). 
\label{eq:mu0c}
\end{eqnarray}
We expand the potential around this point by substituting 
$T=T_0+\delta T$, $X=X_0+\delta T$ and find 
$\delta T/T_0 \sim 1/(aT_0)^2 = {\cal O}(1/(4 \pi^2)^2)$, 
$\delta X/X_0 \sim 1/(aT_0) = {\cal O}(1/(4 \pi^2))$. 
The vacuum energy at this SUSY breaking minimum is vanishing, 
$V=0$, if we tune the parameters as 
\begin{eqnarray}
\tilde{c} &\simeq& \mu_0^2/\sqrt{3 Z_0}+{\cal O}(\mu_0^4), 
\label{eq:vve}
\end{eqnarray}
which is consistent with the above assumption 
$\mu_0^2 = {\cal O}(c)$. 
The leading moduli mass $m_T$ and $F^T$ are the same as 
those in the original KKLT model, while we obtain 
\begin{eqnarray}
F^X &\simeq& \sqrt{3/Z_0}\,m_{3/2} \ = \ {\cal O}(m_{3/2}). 
\label{eq:fx}
\end{eqnarray}

In general, the K\"ahler mixing at the reference point 
(\ref{eq:refpt}) is suppressed 
$K_{T \bar{X}}, K_{TX} \propto X_0 = {\cal O}(c)$ 
satisfying the first two conditions in Eq.~(\ref{eq:mixcd}), in the 
ISS-type model where the VEV of $X$ can be significantly small. 
In such a case, we find general expressions~\cite{Abe:2007yb}, 
\begin{eqnarray}
m_T &\simeq& 
\left. -e^{K/2}K^{T \bar{T}}W_{TT} \right|_{T=T_0,X=X_0}, 
\nonumber \\
F^T &\simeq& \left. 
-\frac{\sqrt{3}K_T}{K_{T \bar{T}}} \left( 
\sqrt{3}+\frac{\sqrt{K^{X \bar{X}}}W_{TX}}{K_T W} 
\right) \frac{|m_{3/2}|^2}{m_T} 
\right|_{T=T_0,X=X_0}, 
\nonumber \\
F^X &\simeq& 
\left. -\sqrt{\frac{3}{K_{X \bar{X}}}}\,
m_{3/2} \right|_{T=T_0,X=X_0}, 
\qquad 
F^C \ \simeq \ \left. C m_{3/2} \right|_{T=T_0,X=X_0}. 
\label{eq:genft}
\end{eqnarray}
This can be also adopted to the case with a sizable 
value of $W_{TX}$ 
under the assumption that the reference point (\ref{eq:refpt}) 
is stable. 

We can generalize the ISS-KKLT model to the case with, e.g., 
$\mu^2(T)=Be^{-bT}$ where the superpotential mixing 
$W_{TX}$ is sizable~\cite{Abe:2007yb}. 
In this case, the perturbation around the reference point 
(\ref{eq:refpt}) becomes unstable if the superpotential 
terms of the modulus sector is KKLT-type. 
We can stabilize the reference point by considering a 
racetrack-type modulus sector (\ref{eq:rt}), i.e., 
\begin{eqnarray}
W &=& Ce^{-cT}-Ae^{-aT}+Be^{-bT}X, 
\label{eq:issrt}
\end{eqnarray}
which we call the ISS-racetrack model. 
In this case, the deviations from the reference point 
(\ref{eq:refpt}), are estimated as 
$\delta T/T_0 \sim bT_0/(aT_0\,cT_0)^2 
= {\cal O}(1/(4 \pi^2)^3)$, 
$\delta X/X_0 \sim (bT_0)^2/(aT_0\,cT_0)^2 
= {\cal O}(1/(4 \pi^2)^2)$. 
Because of the racetrack structure, the modulus mass is 
enhanced as Eq.~(\ref{eq:rtmt}), but unlike Eq.~(\ref{eq:rtft}), 
$F^T$ is estimated by Eq.~(\ref{eq:genft}) as 
\begin{eqnarray}
\frac{F^T}{T+\bar{T}} &\simeq& \frac{b}{(aT_0)(cT_0)}m_{3/2} 
\ = \ {\cal O}(m_{3/2}/4 \pi^2), 
\nonumber
\end{eqnarray}
due to the enhancement factor $b$ from the $T$-$X$ mixing, 
$W_{TX}$ (i.e., the second term in the parenthesis of $F^T$ 
in Eq.~(\ref{eq:genft})). 
Thus the tree-level modulus mediation is comparable to 
the anomaly mediation, in spite of a larger mass hierarchy 
between the modulus and the gravitino mass (\ref{eq:rtmt}).

\end{document}